\documentclass[12pt]{article} 

\usepackage{graphicx}
\title{Expansion of non-neutral, local equilibrium plasma}
\author{A.R.~Karimov$^{1,2}$, A.A. Dementev$^1$\\
$^1$Department of Electrophysical Facilities, \\ National Research Nuclear University MEPhI, \\
Kashirskoye shosse 31, Moscow, 115409, Russia \\
$^2$Institute for High Temperatures, \\
Russian Academy of Sciences,\\
Izhorskaya 13/19, Moscow 127412, Russia}
\date{ }
\begin{document} 
\maketitle	
\newcounter{graf}
\begin{abstract}
The nonlinear dynamics of collisionless non-neutral plasma without external stabilizing factors is considered. 
Time-dependent one-particle distribution functions of a Maxwellian type are obtained.
The influence of initial conditions on the entropy production is discussed.
\end{abstract}
\maketitle

\section{Introduction}

Non-neutral, collisionless plasmas are of interest in natural phenomena (e.g., the studies of solar wind or relaxation of turbulence in the incompressible viscous flows \cite{JD_1998,DN_1999}) and in technical applications (e.g., klystrons, particles accelerators \cite{Dv_1974,anderegg}). The non-neutral media have large self-generated electric fields and they can not exist without an external stabilizing factors. As a rule, non-neutral non-relativistic  plasma is confined by external electrostatic or magnetic traps \cite{DN_1999,anderegg} when  we faced with a quasi-stationary state, which characterized by certain self-generated electric fields. In this case, different types of nonlinear equilibria are possible due to the collective character of interaction between particles and waves. It means that depending on the initial values, the system can evolve in many ways from the initial state far from equilibrium to the state where distribution function has stop varying with time.

One of the possible ways for such evolution is a dynamics via the formation of local equilibrium state formed in the self-generated and external holding fields. It is well known that to obtain exact nonlinear solutions even for a simple plasma system with arbitrary initial/boundary conditions is generally difficult. In order to investigate the possibility of steady states, waves and patterns for such system it would be useful to consider the dynamical properties of physically more simple system but in a fully nonlinear formulation \cite{Levin_2008,Krishan_1990}. In the present paper we are going to study the case in one dimensional geometry for the cloud of the single charged particles without external trapping field. In order to investigate the evolution of such  a system, we shall construct time-dependent Maxwellian distribution functions satisfying the one-dimensional Vlasov-Poisson equations.

\section{Formulation of the kinetic problem}\suppressfloats

We consider the dynamics of  non-neutral, non-relativistic collisionless one-dimensional slab  in electrostatic approximation, where distribution function $ f=f(t, x,v)$ is determined by Vlasov-Poisson equations written in dimensionless form as
\begin{equation} 
\frac{\partial f}{\partial t}+v \frac{\partial f }{\partial x} - {\rm
sgn}(q)\frac{\partial \Phi }{\partial x}\frac{\partial f }{\partial
v}=0\/,
\label{1_nature}
\end{equation}
\begin{equation} 
\frac{{\partial}^{2}\Phi}{\partial x^2}= - {\rm sgn}(q) n= - {\rm sgn}(q)
\int_{-\infty}^{+\infty}f(t,x,v)dv\/,
\label{2_nature}
\end{equation}
where $q$ is the charge of the particles, which form the medium, $\Phi(t,x)$ is the electrostatic potential, which determines the self-generated field of non-neutral plasma. Here we take the initial electron density $n_0$, the values $\omega_b=(4\pi n_0 e^2/m_e)^{1/2}$ and  $\lambda_b=(T_e/4\pi e^2n_0)^{1/2}$, where $m_e$ is the mass of electron and $T_e$ is the temperature of electron,  as the scale of the electron density, lengths and time.  The  velocity is normalized by $\omega_b \lambda_b$, and  potential is normalized by $T_e/e$.

Assume that the system is in the state of local equilibrium. Then
we can restrict our consideration to the partial class of one-particle
distribution functions in the form
\begin{equation} 
f(t,x,v)=\frac{n}{\sqrt{T\pi }}
\exp\left(-\frac{\left(v-V\right)^2}{2}\right)\/,
\label{3_nature}
\end{equation}
where $n=n(t)$, $T=T(t,x)$, $V=V(t,x)$ are unknown functions to be determined. The corresponding potential has the following form:
\begin{equation} 
\Phi ={b}_{0}(t)+{b}_{1}(t)x+\frac{{b}_{2}(t)}{2}x^2\/.
\label{4_nature}
\end{equation}

From (\ref{2_nature}) and (\ref{4_nature}), we get 
\begin{equation} 
b_2(t)= - {\rm sgn}(q) n(t)\/.
\label{5_nature}
\end{equation}
From symmetry consideration
$$\left.\frac{\partial\Phi }{\partial x} \right|_{x=0}=0\/,$$
it follows $b_1(t)\equiv 0$ and we have
\begin{equation} 
\frac{\partial\Phi}{\partial x}= - {\rm sgn}(q) n(t)x
\label{6_nature}
\end{equation}
By inserting (\ref{3_nature}) into (\ref{1_nature}) with
(\ref{6_nature}) we obtain
$$\left[\frac{\dot{n}}{n}+\frac{\dot{T}}{T^2}V^2-\frac{\dot{T}}{2T}-\frac{2V\dot{V}}{T}
+\frac{2nx}{T}V \right]v^0 + \left[\frac{2\dot{V}}{T}
-\frac{2V\dot{T}}{T^2} -\frac{T'}{2T} +\frac{T'V^2}{T^2}-\frac{2V'V}{T} -
\frac{2nx}{T}\right]v^1 +$$
$$+\left[\frac{\dot{T}}{T^2}-\frac{2T'V}{T^2}+\frac{2V'}{T}\right]v^2 +
\frac{T'}{T^2} v^3 =0\/,$$
which is valid for any $v$. So, the coefficients of each power of $v$ should be equaled to zero to yield
\begin{equation} 
\frac{\dot{n}}{n}+\frac{\dot{T}}{T^2}V^2-\frac{\dot{T}}{2T}-
\frac{2V\dot{V}}{T}+\frac{2nx}{T}V=0\/,
\label{8_nature}
\end{equation}
\begin{equation} 
\frac{2\dot{V}}{T}-\frac{2V\dot{T}}{T^2}
-\frac{T'}{2T}+\frac{T'V^2}{T^2}-\frac{2V'V}{T} - \frac{2nx}{T}=0\/,
\label{9_nature}
\end{equation}
\begin{equation} 
\frac{\dot{T}}{T^2}-\frac{2T'V}{T^2}+\frac{2V'}{T}=0\/,
\label{10_nature}
\end{equation}
\begin{equation} 
\frac{T'}{T^2}=0\/.
\label{11_nature}
\end{equation}

As is seen from (\ref{11_nature}), temperature $T$ does not depend on the coordinate $x$. It is only a function of time, and hence (\ref{9_nature}) and (\ref{10_nature}) become
\begin{equation} 
\frac{2\dot{V}}{T}-\frac{2V\dot{T}}{T^2} -\frac{2VV'}{T} - \frac{2n}{T}x
=0\/,
\label{12_nature}
\end{equation}
\begin{equation} 
\frac{\dot{T}}{T}+2V'=0\/.
\label{13_nature}
\end{equation}
Since $\dot{T}/T$ depends only on time, then from (\ref{13_nature})
we get
\begin{equation} 
V(t,x)=-\frac{1}{2}\frac{\dot{T}}{T}x\/.
\label{14_nature}
\end{equation}
Substitution of (\ref{14_nature}) into (\ref{12_nature}) and
(\ref{8_nature}), leads to
\begin{equation} 
\frac{d}{dt}\left(\frac{\dot{T}}{T}
\right)=\frac{1}{2}\left(\frac{\dot{T}}{T} \right)^2 - 2n
\label{15_nature}
\end{equation}
and
\begin{equation} 
\left[\frac{\dot{n}}{n}-\frac{\dot{T}}{2T}
\right]-\frac{\dot{T}}{2T^2}\left[\frac{d}{dt}\left(\frac{\dot{T}}{T}
\right)-\frac{1}{2}\left(\frac{\dot{T}}{T} \right)^2+2n \right]x^2=0\/.
\label{16_nature}
\end{equation}

As is seen from (\ref{16_nature}), the second bracket is equivalent to
(\ref{15_nature}), so the relation (\ref{16_nature}) is satisfied under 
\begin{equation} 
\frac{\dot{n}}{n}=\frac{\dot{T}}{2T}\/.
\label{17_nature}
\end{equation}
This equation has the solution
\begin{equation} 
n=n_0{T}^{1/2}\/,
\label{18_nature}
\end{equation}
For convenience, here we set $C=n_0$ where $C$ is the constant of integration. This makes it possible to eliminate the density $n(t)$ from the equation
(\ref{15_nature}). As a result, we get
\begin{equation} 
\frac{d^2 u}{dt^2}-\frac{1}{2}\left[\frac{du}{dt} \right]^2 +
2n_0 {e}^{u/2}=0\/,
\label{19_nature}
\end{equation}
here, we have used the notation
\begin{equation} 
u=\ln T\/.
\label{20_nature}
\end{equation}

Using (\ref{20_nature}), (\ref{18_nature}) and (\ref{14_nature}),
we express parameters in the distribution (\ref{3_nature}) in explicit form through $u$:
\begin{equation} 
n=n_0e^{u/2}, \hspace{9mm} T= e^u, \hspace{9mm} V=
-\frac{1}{2}\frac{du}{dt}x\/.
\label{a_nature}
\end{equation}
Thus, proceeding from these relations and Eq. (\ref{19_nature}) we can determine the evolution of the system worked out.

\begin{figure}[h]
\begin{center}
\includegraphics[width=11.cm]{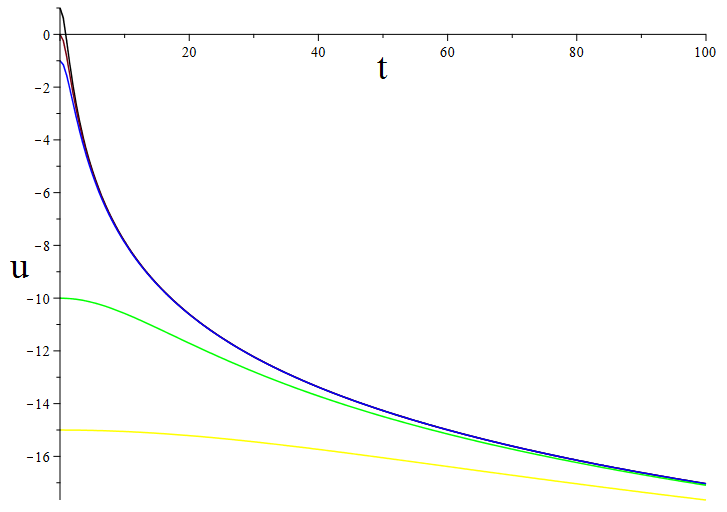}  
\end{center}   
\caption{Solution of equation (18) for $ u'(t=0)=0 $ for $u_0=1$ (black curve), $u_0=0$ (red curve), $u_0=-1 , $ (blue curve), $u_0=-10$ (green curve), $u_0=-15$ (yellow curve) \label{fig_1}}
\end{figure}

\section{Dynamic properties}

The physical properties of non-neutral plasma is completely determined by
dynamical properties of the governing Eq. (\ref{19_nature}). In the absence of the heating or cooling sources we have to put
$$\left.\frac{\partial u }{\partial t}\right|_{t=0}=0\/,$$
that corresponds to the immobile non-neutral medium for the  initial moment of time. Depending on the initial state, there exists a rich variety of solutions of this nonlinear equation. 

As an simple illustration, we consider the influence of initial temperature on the dynamics of the system worked out. Fig. 1 shows such dependence with decreasing temperatures from $u_0=1$ to $u_0=-15$. According to (\ref{a_nature}) the monotonic decrease of $u(t)$ leads to an exponential decrease the density $n(t)$ and temperature $T(t)$ for the non-neutral medium. This process accompanies the growth in the macroscopic velocity $V(t)$ over the course of time. As is seen from these curves, for large times when there is a little influence of the initial conditions and  the curves have the same dependence.

The behavior of this dependence can be determined from the simplified
equation (\ref{19_nature}). For $u(t)< 0$ and $\mid u(t)\mid \ll 1$ 
this equation reduces to
$$\frac{d^2 u}{dt^2}-\frac{1}{2}\left[\frac{du}{dt} \right]^2=0\/,$$
whose solution has the form
$$u(t)=C_{**}- 2\ln(C_* +t)\/,$$
where $C_{**}$ and $C_*$ are some constants. From the comparison of this dependence with the curves in Fig.1, it follows that such behavior is determined by the nonlinearity of the initial problem.

It is interesting to use these results to evaluate the Gibbs entropy
\begin{equation} 
S=-\int f \ln f dvdx
\label{21_nature}
\end{equation}
for the non-neutral plasma slab. Inserting (\ref{3_nature}) into (\ref{21_nature}), we obtain
\begin{equation} 
S=C\int n dx\/,
\label{22_nature}
\end{equation}
where $C=1+\ln(\pi^{1/2}/n_0)$, and consequently the dynamics of  entropy is defined by
\begin{equation} 
\frac{dS}{dt}=\frac{Сn_0 x}{2}\frac{du}{dt} \exp(u/2)=
\frac{C}{2}n(t)V(t)\/.
\label{23_nature}
\end{equation}
This relation  shows that for sufficiently large initial
temperatures the entropy increases monotonically with the expansion of 
non-neutral medium in accordance with the curves in Fig. 1. As seen from these graphs, the cooling of medium leads to decrease in the derivative $\mid du/dt \mid$ that leads to the decrease in the production of entropy. However, with decreasing $\mid du/dt \mid \sim
1/(C_*+t)$ one can see the growth of exponential term in
(\ref{19_nature}). Therefore, it would be interesting to study the question about the temperature evolution for large times in details. 

\section{Conclusion}

The dynamics of non-neutral plasma considered here is special case. 
We again would like to emphasize that the one-dimensional model, the nonlinear dynamics of which we have analyzed here rigorously on mathematical grounds, is still rather artificial because real systems are
essentially higher dimensional and, in the absence of trapping, dissipation. On the other hand, the exact solutions describing possible highly nonlinear final states should be useful as a bench test for new analytical approach. One can expect that some physical variables for real system inherit the general characteristics of the present primitive solution. Namely, this model indicates what may come about in real systems. In particular, from the presented results we may catch a glimpse of the formation of time-dependent structures in higher dimensions governed by the hydrodynamic nonlinearity as a reference to a more realistic dynamical evolution \cite{k97}-\cite{ks_15}.

\end{document}